\documentclass[12pt]{iopart}

\usepackage{graphicx}
\usepackage{amssymb}
\usepackage[ansinew]{inputenc}
\usepackage{epsfig}
\usepackage{latexsym}
\usepackage{color}

\begin{document}
\title{Is keV ion induced pattern formation on Si(001) caused by metal impurities?}

\author{Sven Macko$^1$, Frank Frost$^2$, Bashkim Ziberi$^2$,
Daniel F. F\"orster$^1$, Thomas Michely$^1$}

\address{$^1$ II. Physikalisches Institut, Universit\"at zu K\"oln,
Z\"ulpicher Str. 77, D-50937 Cologne, Germany}

\address{$^2$ Leibniz-Institut für Oberfl\"achenmodifizierung e.~V.,
Permoserstr. 15, D-04318 Leipzig, Germany}

\eads{\mailto{michely@ph2.uni-koeln.de},
\mailto{macko@ph2.uni-koeln.de}}

\date{\today}

\begin{abstract}
We present ion beam erosion experiments performed in ultra high
vacuum using a differentially pumped ion source and taking care
that the ion beam hits the Si(001) sample only. Under these
conditions no ion beam patterns form on Si for angles $\vartheta \leq
45^\circ$ with respect to the global surface normal using 2\,keV
Kr$^+$ and fluences of $ \approx 2 \times 10^{22}\,\rm{ions/m^2}$.
In fact, the ion beam induces a smoothening of preformed patterns.
Simultaneous sputter deposition of stainless steel in this angular
range creates a variety of patterns, similar to those previously
ascribed to clean ion beam induced destabilization of the surface
profile. Only for grazing incidence with $60^\circ \leq \vartheta
\leq 83^\circ$ pronounced ion beam patterns form. It appears that
the angular dependent stability of Si(001) against pattern
formation under clean ion beam erosion conditions is related to
the angular dependence of the sputtering yield, and not primarily to a
curvature dependent yield as invoked frequently in continuum
theory models.
\end{abstract}

\pacs{81.16.Rf, 81.65.Cf, 68.49.Sf, 81.15.Jj, 87.64.Dz}
\submitto{\NT} \noindent{\it Keywords\/}: Si(100), silicon,
surface, krypton, Kr, ion beam, pattern formation, co-deposition,
STM, nanoholes, nanodot, nanoripples, smoothening, impurities, Fe,
iron

\maketitle

\section{Introduction}

Ion beam surface patterning of amorphous and crystalline materials
has attracted considerable interest in the recent years. This
interest is of twofold origin. First it results from the intricate
physics involved in ion beam patterning which has not yet been
entirely uncovered. Second, ion beam nanopatterning is a
relatively cheap and low tech method with a number of potential
applications ranging from an anti-reflection surface finish
\cite{Wilson_82,Flamm_01,Fouckhardt_07} over orienting large
molecules \cite{Chaudhari_01} to nanomagnetism
\cite{Zhang_07,Zhang_08,Fassbender_07,Teichert_09}. As Si is a
prime material of technology and readily available in high purity
and quality, it is not surprising that the ion beam patterning
studies are numerous for Si. At room temperature Si readily
amorphizes during ion exposure \cite{Kotai_94}. The loss of
anisotropy and crystal structure appeared to make it an ideal
material to be described by the continuum theory of ion erosion,
which effectively averages out atomistic details of the processes.
Early theoretical work in the continuum theory approximation
considered the dependence of the sputtering yield $Y$ on the angle
$\theta$ of the ion beam with the respect to the \emph{local}
surface normal as a decisive factor for surface morphological
evolution (\cite{Katardjiev_89,Carter_01} and references therein).
Note that we distinguish here and in the following between the
\emph{local} angle of incidence $\theta$ (measured with respect to
the surface normal of a specific surface element) and the
\emph{global} angle of incidence $\vartheta$ (measured with
respect to the normal of the average surface plane). Since the
seminal publication of Bradley and Harper \cite{BradleyHarper_88}
the dependence of the sputtering on surface curvature was
considered to be the key for pattern formation. While there are
certain conditions, where ion beam erosion of Si does not cause
pattern formation \cite{Madi_08,Ziberi_09}, the overwhelming
number of investigations find pattern formation on Si in a large
parameter space. One class of prototypical patterns are dot or
hole patterns observed for normal incidence noble gas ion erosion
with energies up to a few keV and at temperatures in the
amorphization regime \cite{Ziberi_09,
Gago_01,Gago_06,Fan_05,Fan_06,Ziberi_05b}. Another class of
exemplary patterns are ripple ones with the ripple wave vector
$\vec{k}$ parallel to the ion beam azimuth. Quite some
observations here refer to elevated temperatures in the
crystalline regime \cite{Erlebacher_99,Erlebacher_00,Brown_05}.
However, also at room temperature in the amorphous regime ripple
patterns were observed
\cite{Gago_02,Habenicht_02,Keller_08,Keller_09}, the most regular
ones for slight off-normal conditions with $\vartheta \approx
15^\circ$ using noble gas ions with energies up to a few keV
\cite{Ziberi_09, Ziberi_05,Ziberi_06,Ziberi_06b}. The variety and
complexity of observed patterns stimulated the development of
continuum theories
\cite{Makeev_02,Facsko_04,Castro_05,Vogel_07,Chan_07} extending
the concept of the curvature dependent yield as prime
destabilization mechanism.

Co-deposition of trace amounts of foreign species during ion
erosion has been found already long time ago to give rise to
microstructure formation
\cite{Wehner_71,Punzel_72,Robinson_82,Rossnagel_82}. It is
currently used as a tool for surface texturing
\cite{Tanemura_04,Hofsaess_09}. Recently it became obvious that
impurities may influence pattern formation unintentionally. Mo
co-sputtered from sample clips during ion erosion was found to
foster dot formation at normal incidence
\cite{Ozaydin_05,OzaydinInce_09,Teichert_06}. The ion flux and
fluence were found to affect the amount of Mo and Fe deposited on
the sample from the ion source and thereby to change normal
incidence patterns \cite{SanchezGarcia_08,SanchezGarcia_09}.

In this situation the question arises, whether there is a
hidden chemical or impurity factor in pattern formation on Si.
Such a largely disregarded attribute could make it impossible for
theory to come up with an adequate material parameter based
description of pattern formation. We therefore conducted an
erosion study under clean surface science conditions, using ion
beam parameters for which we expected pattern formation to take
place.

\section{Experimental}

The experiments were performed in a variable temperature scanning
tunneling microscopy (STM) apparatus \cite{Polop_07} with a base
pressure $< 6 \times 10^{-11}$~mbar equipped with a differentially
pumped ion source, a low energy electron diffraction (LEED)
system, and a load-lock for Si sample transfer. For erosion the
samples were exposed to a 2\,keV Kr$^+$ fine focus ion beam with
full width at half maximum $\approx 1$\,mm scanned over a sample
area of about 4\,mm\,$\times$\,4\,mm resulting in a time averaged
ion flux of $5 \times 10^{16}\,\rm{ions\,m^{-2}\,s^{-1}}$ at
300\,K at angles $0^\circ \leq \vartheta \leq 83^\circ$. The
fluence was $F \approx 2 \times 10^{22}\,\rm{ions\,m^{-2}}$. Due
to differential pumping during ion exposure the working pressure
was below $9 \times 10^{-8}$\,mbar. The ion current onto the
sample was controlled with a Faraday cup, which could be moved
precisely into the sample position. After ion exposure the
pressure dropped quickly into the $10^{-11}$\,mbar range and
imaging by STM was performed subsequently. For the co-sputter
deposition experiments a piece of target material was mounted
vertically on the sample. The ion beam was impinging at $\vartheta
= 30^\circ$ onto the sample surface and with an angle
$\vartheta_{dep} = 60^\circ$ onto the sputter target. In the
co-sputter deposition experiments the fluence was only $F \approx
5 \times 10^{21}\,\rm{ions\,m^{-2}}$, but still sufficient to
guarantee pattern development. Again the samples were analyzed in
situ by STM and subsequently analyzed ex situ by atomic force
microscopy (AFM), secondary electron microscopy (SEM), a compact
phase-shifting interferometer and secondary ion mass spectrometry
(SIMS). For SIMS we used 250\,eV O$_{2}^{+}$ for depth profiling.
Quantitative image analysis was conducted by WSxM \cite{Horcas_07}
and differential sputter yield and energy distribution
calculations were performed with TRIM.SP \cite{Eckstein_94}.

\section{Angle dependent smoothening and pattern formation on Si(001)}

\begin{figure}[t, b]
\centerline{\includegraphics[width=17.8cm]{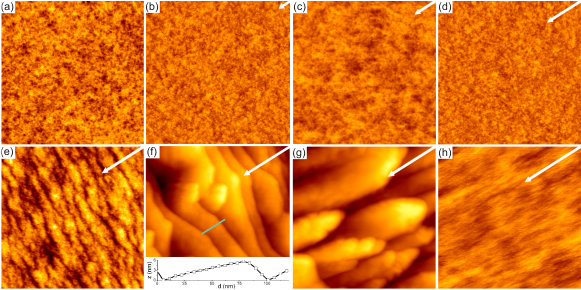}}
\caption{STM topographs of Si(100) after a fluence of $2 \approx
10^{22}\rm{ions\,m^{-2}}$ 2\,keV $Kr^{+}$ at 300\,K. The angle of
incidence $\vartheta$ with respect to the surface normal is (a)
$0^\circ$, (b) $15^\circ$, (c) $26^\circ$, (d) $45^\circ$, (e)
$60^\circ$, (f) $75^\circ$, (g) $79^\circ$ and (h) $83^\circ$. The
image size is for (a)-(d) and (h) 316\,nm\,$\times$\,316\,nm and
in (e)-(g) 625\,nm\,$\times$\,625\,nm. The white arrows in (b)-(h)
indicate the ion beam azimuth. The corrugation $\Delta z$ is 3\,nm
in (a)-(e) and (h) and 30\,nm in (f) and (g). Inset in (f): height
profile along line indicated in (f).} \label{figure1}
\end{figure}

The $\vartheta$-dependence of the morphology after ion bombardment
is shown in Fig.~\ref{figure1}. Unexpectedly, for $0^\circ \leq
\vartheta \leq 45^\circ$ no patterns form and the root mean square
roughness $\sigma$ remains very low, $\sigma \approx 0.2$\,nm
[compare Figs.~\ref{figure1}(a) - \ref{figure1}(d)]. For
$\vartheta = 60^\circ$ as shown in Fig.~\ref{figure1}(e) ripples
with a small amplitude and wave vector $\vec{k}$ parallel to the
ion beam azimuth developed. From the power spectral density of
large topographs we obtain an average wavelength of $\lambda
\approx 46$\,nm. Still the surface does not destabilize to any
significant extent: $\sigma$ is just $0.5$\,nm after removal of
the order of $1\,\rm{\mu m}$ of material. Pronounced pattern
formation takes place in a narrow angular range $75^\circ \leq
\vartheta \leq 79^\circ$, just around the angular range of maximum
sputter yield $Y(\vartheta)$. At $\vartheta =75^\circ$ represented
by Fig.~\ref{figure1}(f) we find a pronounced sawtooth profile
ripple pattern with $\sigma = 6.4$\,nm. The sawtooth profile
displays extended facets forming a small angle of $\approx
7^\circ$ with respect to the global surface plane resulting in an
angle $\theta \approx 82^\circ$ of the ion beam with respect to
the \emph{local} surface normal [compare inset of
\ref{figure1}(f)]. The smaller facets form an apparent angle of
$\approx 19^\circ$ with respect to the surface plane, resulting in
an apparent angle $\theta \approx 56^\circ$ of the ion beam with
respect to the \emph{local} surface normal. Due to convolution of
the surface profile with the STM tip of finite sharpness the
apparent $\theta \approx 56^\circ$ is just an upper bound to the
true angle of the ion beam with respect to the local surface
normal, which is likely to be much lower. At $\vartheta =
79^\circ$ represented by Fig.~\ref{figure1}(g) instead of ripples
we find roof tile structures \cite{Hofsaess_09}. The structures
are now elongated along the ion beam and if one would like to
assign a $\vec{k}$ to them (which is not justified), it would now
be normal rather than parallel to the ion beam azimuth. The
roughness is with $\sigma = 5.3$\,nm similar to the $\vartheta =
75^\circ$ case. Note also that the facet structure of the roof
tiles in Fig.~\ref{figure1}(g) has similarities, but also
discrepancies compared to Fig.~\ref{figure1}(f). Also the roof
tiles display extended facets parallel to the ion beam direction
with a small angle $\approx 3^\circ$ with respect to the global
surface plane resulting in $\theta \approx 82^\circ$. The facets
normal to the ion beam direction visible in Fig.~\ref{figure1}(f)
became arrow tips. The angle of the arrow tips with respect to the
average surface plane is $\approx 19^\circ$ limited again by the
surface profile -- tip convolution. In Fig.~\ref{figure1}(h) for
$\vartheta = 83^\circ$ the surface is extremely smooth again with
$\sigma = 0.2$\,nm. A faint ripple pattern with $\vec{k}$ normal
to the ion beam azimuth is visible; however, the amplitude of the
pattern is marginal. If forced to define a critical angle of
ripple rotation we would set it to $\vartheta_c = 77^\circ$, in
reasonable agreement with previous work \cite{Hofsaess_09} finding
a $\vartheta_c = 80^\circ$ for 5\,keV Xe$^+$ ion erosion of Si.

\begin{figure}[t, b]
\centerline{\includegraphics[width=17.8cm]{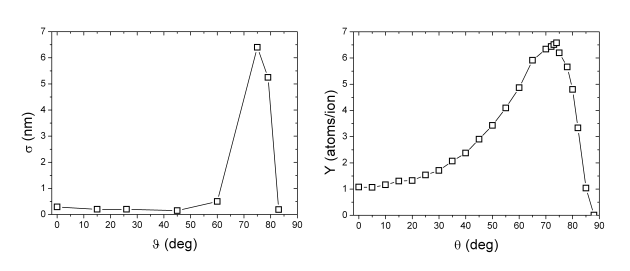}}
\caption{(a) Roughness $\sigma$ as a function of the angle of
incidence $\vartheta$ for the experiments represented by
Fig.~\ref{figure1}. (b) Sputtering yield $Y(\theta)$ as a function
of the angle $\theta$ of the ion beam with respect to the
\emph{local} surface normal for 2\,keV Kr$^+$ on Si as calculated
by TRIM.SP \cite{Eckstein_94}. Lines to guide the eye.}
\label{figure2}\label{theta}
\end{figure}

Figure~\ref{figure2}(a) summarizes the evolution of $\sigma$
already discussed during the presentation of the STM data above.
The absence of roughening for $\vartheta \leq 45^\circ$ is no
accident, but due to an inherent smoothening action of the ion
beam. To support this idea after STM imaging of the rough
$\vartheta = 75^\circ$ ripple pattern with $\sigma = 6.4$\,nm, we
exposed it to a fluence of $\approx 5 \times
10^{21}\,\rm{ions\,m^{-2}}$ of 4\,keV Kr$^+$ at normal incidence
($\vartheta = 0^\circ$) and halved thereby the roughness.

In order to discuss the mechanism of pattern formation for
$60^\circ \leq \vartheta \leq 79^\circ$ we consider the angular
dependence of the sputter yield $Y$ on the local angle of
incidence $\theta$ as calculated by TRIM.SP and shown in
Fig.~\ref{theta}(b).  Evidently, this distinction is only
relevant, when the surface is not flat. $Y(\theta)$ displays a
broad local minimum for $\theta \approx 0^\circ$, an absolute
minimum for $\theta = 90^\circ$ with $Y(90^\circ) = 0$ and a
strong maximum around $\theta_P \approx 75^\circ$ (see also
\cite{Ducommun_1975}). It is obvious, that pronounced patterns
develop only for global angles of incidence $\vartheta$ close to
the maximum of $Y$, i.e. where the erosion rate is largest and
where the erosion rate depends strongly on $\vartheta$. This
points to the application of the deterministic approach of erosion
profile evolution, where the sputtering yield is assumed to be a
function only of the local surface gradient
\cite{Carter_73,Carter_77,Katardjiev_89,Carter_01}.

If the sputter yield depends only on the local angle of incidence
$\theta$, a smooth surface eroded under a global angle $\vartheta$
can become unstable, if fluctuations are present which give rise
to locally varying sputtering yields. An analysis of surface
topography evolution assuming that the sputter yield only depends
on the local incidence angle $\theta$ shows that, due to
fluctuations, an initially flat surface can decompose into a
faceted profile which displays facets where the sputter yield is a
minimum or maximum \cite{Carter_73,Carter_77}. For these facets
which predominate during ion erosion the local ion incidence
angles $\theta$ are given by $0^\circ$, $\pm \theta_P$, or
$90^\circ$, respectively. It should be noted, that $90^\circ$
facets can only form if in the initial surface profile ion
incidence angles $\theta > \theta_P$ are already existing
\cite{Carter_73,Wei_09}. For surfaces with small height
fluctuation this condition is only achieved if $\vartheta$ is
close to $75^\circ$.

As the average surface orientation is maintained during erosion,
in the angular range $\vartheta \approx 75^\circ-80^\circ$ the
surface profile decomposes into patches with local ion impingement
angles $\theta$ close to the minima of $Y(\theta)$, i.~e. into
facets $1$ and $2$ with $\theta_1 < \vartheta < \theta_2$.
Considering the example represented by Fig.~\ref{figure1}(f) with
$\vartheta = 75^\circ$ the flat surface decomposes into a sawtooth
profile with facets forming a large angle with the ion beam
$\theta_1 \leq 54^\circ$ and facets nearly parallel to the ion
beam ($\theta_2 \approx 82^\circ$). For both $\theta_1$ and
$\theta_2$ the yield $Y$ is much smaller than for $\theta =
75^\circ$ [compare Fig.~\ref{theta}(b)]. It is evident that such a
sawtooth profile considerably reduces the global erosion rate.
However, the observed local angles $\theta = 54^\circ$ and $\theta
= 82^\circ$ do not match the theoretical predictions of $\theta =
0^\circ$ and $\theta = 90^\circ$. For the steep facet we attribute
this difference largely to inability of our STM tip to measure the
proper angle. For the facet with the small slope the difference of
the measured $82^\circ$ and predicted $90^\circ$ can not rely on a
measurement problem. The measured difference may have the
following reasons: (i) The fluence used was too low to allow the
surface to reach the dynamic equilibrium and the ensuing facets;
(ii) There is a significant uncertainty in the TRIM.SP
calculations of $Y(\theta)$ for very grazing angles due to the
fast change of the yield with $\theta$ and its sensitive
dependence on surface structure. The yield at $\theta = 82^\circ$
might thus be already close to zero, such that the driving force
(minimization of erosion rate) has largely vanished and a further
change of the facet angle is kinetically frozen.

In contrast, for near normal ion incidence angles $\vartheta \sim
0^\circ$, $\theta < \theta_P$ and surfaces with small height
fluctuations are stable because profiles with $\theta = 0^\circ$
are steady state surface configurations
\cite{Carter_73,Carter_77}. Furthermore, as hillocks are eroded
faster than valleys (contrary to the Bradley-Harper theory
\cite{BradleyHarper_88}) the ion beam induces a smoothening of
rough surfaces in this angular range as observed by us
experimentally (compare also Fig. 4 of \cite{Carter_01} and
\cite{Carter_92}).

For the case of room temperature ion beam erosion of Si with
2\,keV Kr$^+$, the angular dependence of $Y(\theta)$ appears to
explain the ranges of stability observed as well as the
orientation of the pattern facets formed in the unstable regime.
So far we left open, which kind of fluctuations might initiate the
faceting of the surface for $\vartheta \approx \theta_P$. While
the stochasticity of the ion impacts gives rise to fluctuations by
itself, it may be that other effects contribute to fluctuations,
e.g. a curvature dependence of the sputtering yield. However, the
pattern formation scenario we observe here can hardly be
reconciled with models based on Bradley-Harper theory. A curvature
dependent yield as destabilization mechanism neither explains the
extended angular range of stability nor the rather abrupt
angle-dependent transitions from smooth surfaces to faceted
patterns and back to smooth surfaces, i.e. abrupt transitions from
surfaces with zero curvature to ones with curvature singularities.
By invoking the angular dependence of $Y(\theta)$ to explain
surface destabilization, no predictions related to time scales
necessary for pattern evolution nor to characteristic length
scales in the pattern are possible. We therefore consider the
angular dependence of $Y(\theta)$ only as one important element in
a theory of pattern formation on Si. Secondary effects - e.g.
surface diffusion, ballistic drift, viscous flow, etc. - must
probably be considered to be responsible for selection of
characteristic scales.

With ion energies of the order of 1\,keV and for noble gases like
Ar$^+$, Kr$^+$ or Xe$^+$ and angles $\vartheta \leq 30^\circ$ a
large variety of dot and ripple patterns was obtained by us
\cite{Ziberi_09,Ziberi_05,Ziberi_06,Ziberi_06b} and other groups
\cite{Gago_01,Gago_06,Fan_05,Fan_06,Gago_02}. We consider these
results to be at variance with the present findings. It was
noticed recently that the relative concentrations of Mo and Fe
emerging from the ion source during sputtering influence the
pattern appearance at $\vartheta = 0^\circ$
\cite{SanchezGarcia_08,SanchezGarcia_09}. However, whether
patterns evolve at all without impurities was not considered. Our
observations are consistent with the observation of Mo-seeding at
$\vartheta = 0^\circ$ \cite{Ozaydin_05,OzaydinInce_09,Teichert_06}
and the absence of patterns without Mo-seeding
\cite{Ozaydin_05,OzaydinInce_09}. We claim here that patterns are
entirely absent without intentional or unintentional co-deposition
of impurities after noble gas bombardment on Si(001) with energies
of the order of 1\,keV and for $\vartheta \leq45^\circ$. Below we
provide additional evidence for this statement.

\section{Angle dependent pattern formation on Si(001) with co-sputter deposition}

\begin{figure}[t, b]
\centerline{\includegraphics[width=17.8cm]{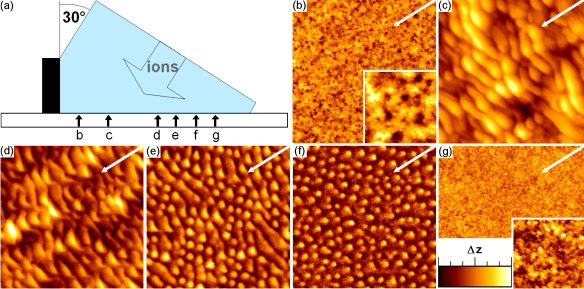}}
\caption{(a) Schematic sketch of co-sputter deposition geometry.
The recording positions of the STM topographs shown in (b)-(g) are
indicated. (b)-(g) STM topographs of Si(001) after a fluence of
$\approx 5 \times 10^{21} \rm{ions\,m^{-2}}$ 2\,keV $Kr^{+}$ at
300\,K at $\vartheta = 30^\circ$. The image width is always
625\,nm; the inset size in (b) and (g) is 100\,nm $\times$ 100\,nm
The ion beam azimuth is indicated by a white arrow. The
corrugation $\Delta z$ is (b) 2\, nm, (c) 45\,nm, (d) 20\,nm, (e)
8\,nm, (f) 4\,nm and (g) 2\,nm.} \label{figure3}
\end{figure}

To substantiate our claim that impurities resulting from the
sputtering process are in fact responsible for the great number of
patterns resulting for ion erosion with $\vartheta \leq 45^\circ$
we performed dedicated co-sputter deposition experiments. As
sketched in Fig.~\ref{figure3}(a) the ion beam was impinging at
$\vartheta = 30^\circ$ onto the sample surface, i.~e. at an angle
where no pattern formation is expected. Additionally the ion beam
hits a piece of stainless steel (Fe 84\% and Cr 13\%) at an angle
$\vartheta_{dep} = 60^\circ$, from which some material is sputter
deposited onto the eroding Si surface. This setup is similar to
the arrangement used for surfactant sputtering \cite{Hofsaess_09}.
The STM topographs were taken along a line normal to the center of
the steel plate. As obvious from Figs.~\ref{figure3}(b)-(g) the
resulting morphologies strongly depend on the normal distance $x$
to the stainless steel plate. Fig.~\ref{figure3}(b) displays a
relatively smooth surface with small hole structures as
highlighted by the inset (compare also \cite{SanchezGarcia_08}).
With increasing $x$ the roughness $\sigma$ shoots up beyond 10\,nm
and ripples with $\vec{k}$ parallel to the ion beam azimuth form
[compare Figs.~\ref{figure3}(c) and (d)]. Upon further increase of
$x$ the roughness $\sigma$ decreases again and the ripple pattern
transforms [Fig.~\ref{figure3}(e)] to a dot pattern
[Fig.~\ref{figure3}(f)]. Eventually, for the largest distance
measured $\sigma$ is comparable to a situation without co-sputter
deposition and patterns are absent [compare insets of
Figs.~\ref{figure3}(b) and (g)].

Not every co-sputtered material induces patterns. We used the same set-up with a
plate of highly oriented pyrolytic graphite instead of stainless steel.
The entire Si wafer remained smooth with a roughness $\sigma \approx 0.2$\,nm
and no patterns formed.

\begin{figure}[t, b]
\centerline{\includegraphics[width=8.7cm]{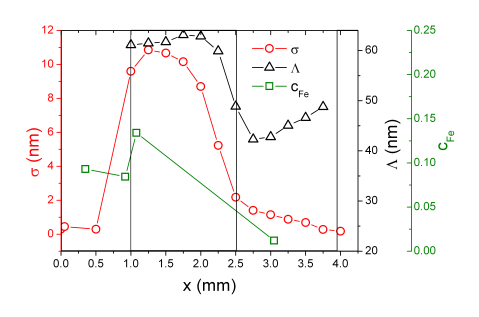}}
\caption{Roughness $\sigma$ (open dots, left $y$-axis) and feature
separation $\Lambda$ (open triangles, right inner $y$-axis) of
morphologies obtained in the co-sputter deposition visualized in
Fig.~\ref{figure3} as a function of the distance $x$ from the
stainless steel plate. The Fe concentration $c_{\rm{Fe}}$
normalized to the Si concentration as function of $x$ is shown on
the outer right $y$-axis (open squares). Lines to guide the eye.}
\label{figure4}
\end{figure}

Figure~\ref{figure4} displays the roughness $\sigma$ (open dots,
left $y$-axis), the feature separation $\Lambda$ (open triangles,
right inner $y$-axis) and the Fe concentration $c_{\rm{Fe}}$ (open
squares, right outer $y$-axis) for the patterns on the Si wafer as
function of $x$. Schematically, four pattern ranges are
distinguished in Fig.~\ref{figure4}: nanoholes with $x =
0\,\rm{mm} - 1.0\,\rm{mm}$, ripples with $x = 1.0\,\rm{mm} -
2.5\,\rm{mm}$, dots with $x = 2.5\,\rm{mm} - 3.8\,\rm{mm}$ and a
smooth surface for $x > 3.8\,\rm{mm}$. The roughness displays a
pronounced maximum in the ripple range with $\sigma \approx
10$\,nm, fades gradually away in the dot range with $\sigma <
2$\,nm and is very low in the nanohole and smooth surface range
with $\sigma \approx 0.3$\,nm. The characteristic feature
separation varies unspectacularly from a typical ripple wavelength
$\Lambda \approx 80$\,nm to a typical dot separation with $\Lambda
\approx 50$\,nm. Based on the variety of patterns we expected an
inhomogeneous Fe concentration $c_{\rm{Fe}}$. As Rutherford
backscattering needs a much larger sample spot (a few mm) for
chemical analysis, we decided to perform secondary ion mass
spectrometry. Quantification of our data is based on the tables of
SIMS impurity signals in Si obtained for 8\,keV O$_{2}^{+}$
sputtering \cite{Wilson_91, Wilson_95}. Here we used 250\,eV
O$_{2}^{+}$ to mimimize implantation and mixing and achieve a high
depth resolution. The surface roughness and the potentially
inhomogeneous Fe distribution makes our quantification
problematic, which therefore should be considered with caution. We
measure a maximum of the relative iron concentration $c_{\rm{Fe}}
\approx 0.13$ in the ripple range. Still $c_{\rm{Fe}} \approx
0.02$ is sufficient to induce the dot range. It appears plausible
that due to ion beam induced mixing the co-sputtered Fe forms a
silicide in the surface near layers as it has been found for
normal incidence ion bombardment through an alternating cold
cathode ion source \cite{SanchezGarcia_08}.

\begin{figure}[t, b]
\centerline{\includegraphics[width=8.7cm]{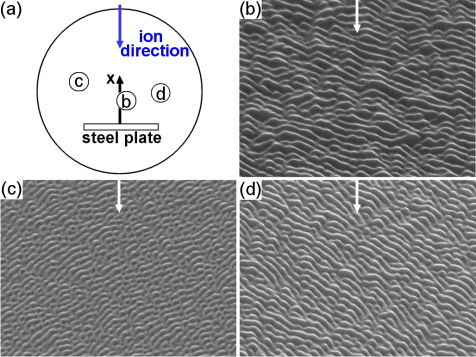}}
\caption{(a) Top view sketch indicating the positions, where
images (b)-(d) were taken by scanning electron microscopy (SEM)
after the erosion experiment visualized by Fig.~\ref{figure3} had
been conducted. $x$-axis indicates the normal to the center of the
steel plate. The direction of e-beam illumination is at $45^\circ$
with respect to the surface normal and the e-beam energy 15\,keV.
The image width in (b)-(d) is 2\,$\mu$m.} \label{figure5}
\end{figure}

We further explored the pattern at locations displaced from the
normal to the center of the stainless steel plate (i.e. left and
right of the x-axis) by scanning electron microscopy.
Figure~\ref{figure5}(a) sketches the locations of the SEM
micrographs displayed in Fig.\ref{figure5}(b)-(d). It is obvious
that the ripple wave vector $\vec{k}$ depends on the position with
respect to the stainless steel plate. Apparently $\vec{k}$ is
parallel to the average azimuthal direction of the atoms sputter
deposited from the stainless steel plate onto the Si wafer.

\begin{figure}[t, b]
\centerline{\includegraphics[width=8.7cm]{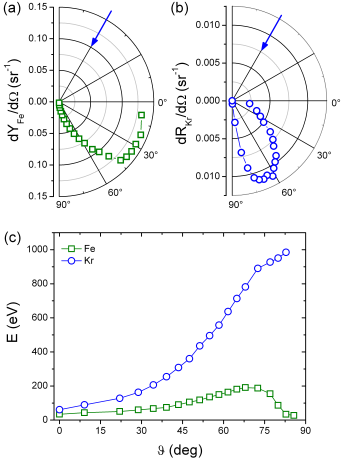}}
\caption{(a) Differential sputtering yield
$\frac{dY_{\rm{Fe}}}{d\Omega}$ and (b) differential backscattering
coefficient $\frac{dR_{\rm{Kr}}}{d\Omega}$ of 2\,keV Kr$^+$
impinging on the steel plate plotted as a function of the polar
emission angle $\vartheta$. The data shown is averaged over the
azimuthal angular range $\varphi \in \left[-30^\circ,
30^\circ\right]$. The arrows indicate the direction of the
impinging Kr$^+$. (c) Energy of the sputtered Fe atoms or
backscattered Kr particles as a function of their polar emission
angle $\vartheta$. The data shown is averaged over the azimuthal
angular range $\varphi \in \left[-30^\circ, 30^\circ\right]$. All
data calculated by TRIM.SP \cite{Eckstein_94}.} \label{polarplots}
\end{figure}

The co-sputter situation is complex and in order to prepare an
understanding of pattern formation in this situation it is
mandatory to consider energy, direction and species of the
particles impinging onto the Si wafer. A given location on the Si wafer in front
of the stainless steel plate receives 2\,keV Kr$^+$ ions,
sputtered Fe atoms (we neglect Cr atoms in the following
discussion for simplicity) and Kr particles reflected from the
stainless plate.  Fig.~\ref{polarplots}(a) displays the
differential Fe sputter yield $\frac{dY_{\rm{Fe}}}{d\Omega}(\vartheta)$
averaged over the azimuthal angular range
$\varphi \in \left[-30^\circ, 30^\circ\right]$ as
calculated by TRIM.SP \cite{Eckstein_94}.
The distribution of the Fe atoms is anisotropic with a
broad peak of emission centered at about $\vartheta
\approx 37^\circ$. As the Si wafer can be imagined horizontally
under the polar diagram, the main flux of sputtered Fe atoms is
directed towards the Si wafer. The total yield $Y_{\rm{Fe}}$
per 2\,keV Kr$^+$ is substantial and amounts to $Y_{\rm{Fe}} = 6.4$.

Fig.~\ref{polarplots}(b) shows the differential backscattering
coefficient $\frac{dR_{\rm{Kr}}}{d\Omega} (\vartheta)$ of the Kr$^+$
particles backscattered from the stainless steel plate
and averaged over the azimuthal angular range
$\varphi \in \left[-30^\circ, 30^\circ\right]$. The
distribution of the backscattered particles is peaked at an exit angle
of $\vartheta \approx 68^\circ$. In average the backscattered
Kr particles will hit the Si wafer closer to the
stainless steel plate than the sputtered Fe atoms, i.e. at a
smaller distance $x$. The total backscattering coefficient
amounts to $R_{\rm{Kr}} = 0.24$.

Fig.~\ref{polarplots}(c) displays the energies $E$ of the
sputtered Fe atoms and of the backscattered Kr particles as a
function of the polar emission angle $\vartheta$. The data plotted
is averaged over the azimuthal angular range $\varphi \in
\left[-30^\circ, 30^\circ\right]$. The energy of the sputtered Fe
atoms is moderate with a maximum of $\approx 210$\,eV for
$\vartheta \approx 74^\circ$. Note that at the most probable angle
of emission of sputtered Fe atoms $\vartheta \approx 37^\circ$ the
energy of the sputtered Fe atoms is only about $65$\,eV in the
presented azimuthal angular range. The average energy of the Fe
atoms sputtered in the azimuthal angular range $\varphi \in
\left[-30^\circ, 30^\circ\right]$ is $112$\,eV, the average energy
of all atoms sputtered in forward direction with $\varphi \in
\left[-90^\circ, 90^\circ\right]$ is $72$\,eV. Though substantial,
the energy of the Fe particles is insufficient to cause
significant sputtering of the Si wafer. The energy of the
backscattered Kr particles increases monotonically with their
emission angle $\vartheta$ and reaches $\approx 1200$\,eV for
$\vartheta \approx 90^\circ$. Note that at the most probable
emission angle of $\vartheta \approx 68^\circ$ the energy of the
emitted particles is about 755\,eV. The backscattered particles
will therefore cause strong erosion of the Si wafer near the steel
plate. The average energy of the backscattered Kr particles in the
presented azimuthal angular range $\varphi \in \left[-30^\circ,
30^\circ\right]$ is $583$\,eV, the average energy of all particles
backscattered towards the Si plate in the angular range $\varphi
\in \left[-90^\circ, 90^\circ\right]$ is $470$\,eV.

\begin{figure}[t, b]
\centerline{\includegraphics[width=8.7cm]{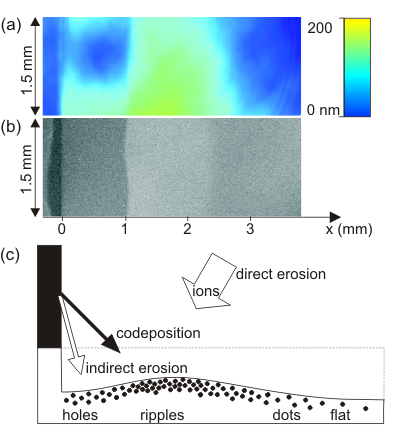}}
\caption{(a) Topography of the Si wafer measured by phase-shifting
interferometry. Center of the long image axis identical with
coordinate $x$. Dark blue contrast for the area with $x < 0$
indicates the position of the steel plate and not height. (b)
Large scale SEM image of the Si wafer. The position of the
stainless steel plate during the erosion experiment is indicated.
(c) Schematic sketch of the erosion situation with co sputter
deposition (see text).} \label{figure7}
\end{figure}

Fig.~\ref{figure7}(a) displays a section of a height topograph of the Si-wafer after co-sputter
deposition measured by phase-shifting interferometry. We attribute the hill in the height profile in the
range $1\,\rm{mm} \leq x \leq 2.5\,\rm{mm}$ with a height of
$\approx 75$\,nm to the significant amount of co-sputtered Fe in
this range. Note that the hill does not consist entirely of Fe -
actually most of the co-sputter deposited Fe has been resputtered
by the Kr$^+$ ion beam. However, the co-sputter deposited Fe
partly shielded the Si from being removed by the direct Kr$^+$ ion
beam resulting in a significantly reduced erosion depth. The
trough next to the stainless steel plate for $0 \leq x \leq
1\,\rm{mm}$ results from the additional sputtering caused by the
energetic component of the Kr particles reflected from the
stainless steel plate. According to mirco phase interference
patterns at the edge of the scanning area of the ion beam the
average erosion depth is $\approx 300$\,nm. The SEM topograph in
Fig.~\ref{figure7}(b) shows a contrast change at the onset of the
co-sputter hill caused by the transition from the smooth nanohole
to the rippled surface. This is also consistent with the pattern
sequence as observed by STM in Fig.~\ref{figure4}. In view of the
significant flux and energy of the reflected Kr particles arriving
in the trough area, there the lower $c_{\rm{Fe}}$ is a consequence
of the enhanced resputtering. Dot patterns form beyond the
co-sputter deposition hill, where the arriving flux of Fe atoms is
lower resulting in a lower $c_{\rm{Fe}}$. Fig.~\ref{figure7}(c)
summarizes the discussion schematically and displays the link
between the arriving particles, the erosion depth and the patterns
formed.

Based on the data the three following conclusions are obtained.
(i) Dot patterns form during 2\,keV Kr$^+$ ion beam erosion with
$\vartheta = 30^\circ$ for moderate co-sputtered Fe concentrations
$c_{\rm{Fe}}$ and in the absence of reflected Kr particles, i.e.
in the absence of a second component of energetic particles
impinging from a different direction onto the substrate. (ii)
Ripple patterns form on the Si substrate for high co-sputtered Fe
concentrations: As the flux and the energy of the reflected Kr
particles arriving in the ripple pattern area $1\,\rm{mm} \leq x
\leq 2.5\,\rm{mm}$ are small, their presence appears not to be
necessary for ripple pattern formation. However Fig.~\ref{figure5}
proves that the direction of impingement of the co-sputtered Fe
atoms determines the orientation of the ripple wave vector
$\vec{k}$. Whether ripple orientation towards the stainless steel
plate is due to the additional directed energy of the arriving Fe
particles or due to the directed attachment of the Fe particles to
elevations (ripple slopes) in their line of sight can not be
resolved by the present experiment. (iii) Erosion by the direct
beam together with strong sputtering from a different direction
(here through the reflected energetic Kr particles) prevents
formation of a pronounced and rough pattern. Although we detect a
significant amount of Fe for $x \leq 1\,\rm{mm}$, the area remains
rather smooth and the nanohole pattern is neither pronounced nor
well ordered.

As a global key result of our experiments we consider the fact
that supply of a second chemical species is mandatory for pattern
formation on Si in a large parameter range, specifically for
$\vartheta \leq 45^\circ$. Whether co-sputter deposition must be
considered also for pattern formation of other materials or other
energy ranges outside the 1\,keV region as a hidden parameter
remains to be investigated. To obtain a deeper understanding
beyond this point and to disentangle the effects of a second
chemical species and of additional energy from a second direction
(at variance with the primary ion beam direction) dedicated
experiments are necessary. It would be useful to design a
situation where the second chemical species is deposited with
thermal energies (e.g. through physical vapor deposition)
simultaneously with a clean eroding ion beam. Such an experiment
might also be able to establish whether the direction of the
impinging co-deposited particles with respect to the ion beam and
the sample surface is of relevance - irrespective of the energy of
the co-deposited particles. As pattern formation is always the
result of the interplay of a destabilizing mechanism and a
stabilizing or smoothening one, the nature of the latter needs to
be explored as well. We suggest to perform co-sputter deposition
or co-evaporation experiments at very low ($\approx 100$\,K) or
elevated temperatures $\approx 450$\,K to test the relevance of
thermal diffusion.

We believe that our experiments are relevant for future
theoretical work. After the decade of continuum theory evolution
there appears now to be a need for (i) a material parameter based
description of pattern formation; (ii) for a theory that takes
into account the absence of pattern formation for a large
parameter space; (iii) a model which includes the effect of
impurities for an explanation of pattern formation.


\section{Conclusions}

Room temperature ion beam erosion of Si through 2\,keV Kr$^+$
shows no pattern formation for $\vartheta \leq 45^\circ$. For
$60^\circ \leq \vartheta \leq 80^\circ$ rough ion beam faceted
ripple and roof tile patterns are formed. It is speculated that
patterns observed for $\vartheta \leq 45^\circ$ are the result of
unintentional co-sputter deposition of metallic impurities
resulting from the use of broad beam ion sources with a divergent beam.
The angular ranges of stability and instability as
well as the observed faceting can be understood on the basis of
the $\theta$-dependence of the sputtering yield.

Intentional co-sputter deposition of stainless steel from a plate
hit together with the Si sample by the ion beam leads to a complex
sequence of patterns in dependence of the distance from steel
plate. These patterns form at angles $\vartheta$, where no
patterns result for clean ion erosion. Nano hole, ripple and dot
patterns are formed. Nano hole patterns are observed under the
influence of strong additional sputtering by ions reflected from
the steel plate and of co-sputtered steel. Ripple and dot patterns
are formed in a distance range where reflected ions are largely
absent. The ripple wave vector of the ripple patterns is found to
be aligned to the direction of the impinging steel atoms. The
formation of ripple and dot patterns is distinguished by the
concentration of co-sputtered material. Dots contain much less
co-sputtered material than ripples.

We believe that our results are relevant for future research in
ion beam pattern formation on Si. They will foster additional
experimental work to uncover the mechanisms of pattern formation
due to the simultaneous deposition of a second chemical species.
They also point to the need for well controlled erosion
experiments and chemical analysis of the eroded surfaces to rule
out impurity effects. Lastly, we hope they stimulate theoretical
work, reconsidering the effect of the slope dependent sputtering
yield for pattern formation and investigating the effects of
impurities on pattern formation.

\ack

We thank Dietmar Hirsch for the SEM and SIMS measurements, Agnes
Mill for optical micro phase interference measurements, Bernd
Rauschenbach for support and useful discussions and Hubert Gnaser
for his advice related to quantification of the SIMS data. We
gratefully acknowledge the support by Deutsche
Forschungsgemeinschaft through Forschergruppe 845.

\section*{References}

\providecommand{\newblock}{}

\end{document}